\documentstyle[psfig,11pt]{article}
\newcommand{\newsection}[1]{
\vspace{10mm}
\pagebreak[3]
\addtocounter{section}{1}
\setcounter{equation}{0}
\setcounter{subsection}{0}
\setcounter{footnote}{0}
\addcontentsline{toc}{section}{\protect
\numberline{\arabic{section}}{{\rm #1}}}
\begin{center}
{\large \sc \thesection. #1}
\end{center}
\nopagebreak
\medskip
\nopagebreak}

\newlength{\extraspace}
\newlength{\extraspaces}
\newcommand{\be}{\begin{equation}
\addtolength{\abovedisplayskip}{\extraspaces}
\addtolength{\belowdisplayskip}{\extraspaces}
\addtolength{\abovedisplayshortskip}{\extraspace}
\addtolength{\belowdisplayshortskip}{\extraspace}}
\newcommand{\ee}{\end{equation}}

\newcommand{\half}{\frac{1}{2}}
\newcommand{\w}{\omega}
\newcommand{\k}{\kappa}

\begin{document}
\begin{flushright}
May 1996\\
ITFA-96-14
\end{flushright}
\thispagestyle{empty}

\begin{center}
{\Large\sc Backreaction on Moving Mirrors\\[7mm] 
and Black Hole Radiation.}\\[22mm] 
{\large \sc  Niels Tuning and Herman Verlinde${}^+$}\\[4mm]
{\it Institute for Theoretical Physics}\\[1mm]
{\it University of Amsterdam}\\[1mm] 
{\it 1018 XE Amsterdam} \\[1mm]
and\\[1mm]
{\it ${}^+$ Joseph Henry Laboratories}\\[1mm]
{\it Princeton University}\\[1mm] 
{\it Princeton, NJ 08544}\\[22mm]

{\sc Abstract}
\end{center}
We compute the effect of quantum mechanical backreaction on the spectrum 
of radiation in a dynamical moving mirror model, mimicing the effect of a 
gravitational collapse geometry. Our method is based on the use 
of a combined WKB and saddle-point approximation to implement energy 
conservation in the calculation of the Bogolyubov coefficients, in which
we assume that the mirror particle has finite mass $m$.
We compute the temperature of the produced radiation as a function
of time and find that after a relatively short time, the temperature
is reduced by a factor $1/2$ relative to the standard result.
We comment on the application of this method to two-dimensional 
dilaton gravity with a reflecting boundary, and conclude 
that the WKB approximation quickly breaks down due to the appearance of 
naked singularities and/or white hole space-times for the relevant 
WKB-trajectories.

\newpage

\newsection{Introduction}

Hawking discovered in 1975 that black holes emit particles due to quantum
effects. The emission spectrum was found to be identical to that of a body 
with temperature  $T=\frac{\k}{2\pi}$, where $\k={1\over 4M}$ equals the 
surface gravity at the black hole horizon \cite{haw}. 
As a result of efforts to understand this effect, Unruh
then showed that the Minkowski vacuum appears to be a thermal
state at temperature $T=\frac{a}{2\pi}$ according to an observer with
acceleration $a$,  \cite{unr}. 
Davies and Fulling subsequently developed 
a simple {\em ``moving mirror''} model, which provides a very 
useful analogy to the black hole situation \cite{dav&ful,bir&dav}. 
By choosing suitable trajectories, 
the moving mirror mimics many of the features of black hole radiance, with
the difference that the quantum fields propagate in flat Minkowski
space instead of complicated geometries with high curvatures. The
trajectory of the moving mirror represents the origin of the
coordinate system of the black hole geometry and both systems display
event horizons, which is essential to get thermal out-going radiation.

The standard derivation of the Hawking effect is based on a semi-classical 
approximation in which one works on a fixed back ground geometry
and in which interactions between the infalling
matter and the out-going virtual particles are ignored. 
As a consequence, a number of basic physical principles are violated 
in this procedure, such as energy conservation and unitarity.
Backreaction effects however could give corrections to the standard
semi-classical result. In this paper we will treat backreaction 
effects in the simple model of the moving mirror, due to the reflection 
of quantum fields off the boundary particle. We will use a WKB-type
approximation.\footnote{For recent related papers see 
\cite{car&wi,ch&ver,kr&wil}.} Contrary to common expectations, we
find that the backreaction has substantial, measurable effects on
the (temperature of) the out-going radiation.

We will begin with a short review the calculation of the out-going 
radiation of 
a mirror with a specified trajectory, which corresponds to a fixed
background geometry in the black hole case. We do this by the method
of saddlepoints, and find the same result as was originally derived
in \cite{dav&ful}. Then we take the rest mass of the mirror {\em
finite} and calculate classically the back reaction effect of the reflection
of massless  scalar fields off the boundary. Again by using a
saddlepoint method we obtain the out-going radiation, which will again be 
thermal radiation, but this time with half the temperature. 

We then comment on the application of the same method to 
two-dimensional dilaton gravity with a reflecting boundary.
Here we find that the WKB approximation 
quickly breaks down because the relevant WKB trajectories 
that should determine the out-going spectrum necessarily contain 
naked singularities and/or white hole space-times.

\begin{figure}[t]
\begin{center}
\hspace{-.5in}
\psfig{figure=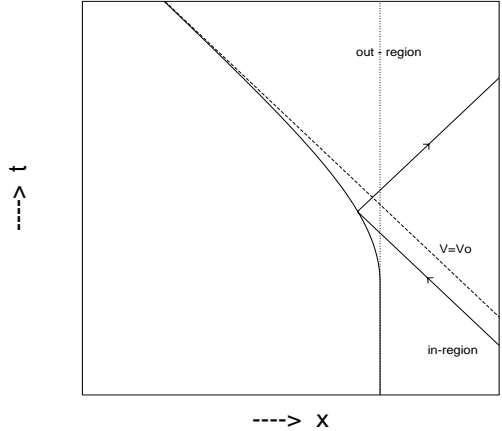,width=2.70in,height=2.25in}
\end{center}
\vspace{.7in}
\caption{{\footnotesize The mirror trajectory $z(t)$ is asymptotic to
the null ray $v=v_0$.}}
\end{figure}

\pagebreak

\newsection{Fixed Mirror Trajectory}

\noindent

Consider a reflecting boundary in 1+1 dimensional Minkowski space, following 
the trajectory 
\be x=z(t).  
\ee
and a massless scalarfield, satisfying the Klein-Gordon equation 
\be
\Box\phi=\partial_u\partial_v\phi=0 
\ee 
with $u = x+t$ and $v=x-t$. We impose the reflecting boundary condition that
the field has to vanish on the boundary, 
\be 
\phi(t,z(t))=0.  
\ee 
The mirror trajectory that mimics the effect of the gravitational
collapse geometry is given by 
\be 
\label{trajectory}
z(t)=-t-\frac{1}{\k}e^{-2\k (t-v_0)}+v_0 
\ee
where $\k$ is a constant (corresponding to the surface gravity at the
horizon)  and $v=v_0$ is the asymptote. The velocity
of the mirror point approaches the speed of light exponentially fast
\be
\label{velocity}
{\rm v}(t) \equiv \frac{d z(t)}{dt} = -1+ 2 e^{-2\k (t-v_0)}
\ee
Left-moving null 
rays after $v=v_0$ do not intersect the mirror trajectory. In the black hole
analogy, these correspond to light-like trajectories that enter into the
black hole region.

Now following the standard procedure to compute the particle spectrum
emitted by this moving mirror, we consider the process 
of a mono-chromatic in-wave $\phi^{in}_{\w}\sim e^{-i\w v}$ 
from past null infinity ${\cal I}^-$ that is being reflected at the boundary. 
A late out-going null ray $u = \bar{u}$ intersects the mirror at
time $\bar{t}=\half(\bar{u}+v_0)$. Thus, given the trajectory $z(t)$,
we get that a reflected light-particle along $u=\bar{u}$ originated
from an incoming particle along the  null-ray $v=\bar{v}$
with
$$
\bar{v}= p(\bar{u})
$$
where
\be
p(\bar{u})= v_0-\frac{1}{\kappa}e^{-\k(\bar{u}-v_0)}.
\ee
The monochromatic in-wave thus corresponds to a reflected out-going wave
of the form
$$
\phi^{out}_{\w}(u) \sim e^{-i\w p(u)}
$$ 
Note, however, that this reflected wave represents only half of the final
wave-function, since the other half at $v>v_0$ will never reach the mirror
and will thus disappear into the left asymptotic region corresponding to
the black hole horizon.

The Bogolyubov transformations (see also \cite{car&wi,bir&dav}) give
the relations between the different expansions $\phi^{out}_{\w'}$ and
$\phi^{in}_{\w}$ of the scalar field. The Bogolyubov coefficients are
computed by evaluating the overlap between the two-types of wave modes
\be
\alpha_{\w\w'}=(\phi_{\w'},\phi_{\w}),\hspace{20pt}
\beta_{\w\w'}=-(\phi_{\w'},\phi^*_{\w}).
\ee
One obtains
$$
\alpha_{\w\w'}= \frac{i}{2\pi}\sqrt{\frac{\w'}{\w}}\int \! du\,
e^{i\w'u-i\w p(u)}
$$
\be \beta_{\w\w'}= -\frac{i}{2\pi}\sqrt{\frac{\w'}{\w}}\int \! du\,
e^{i\w' u +i\w p(u)}.  
\ee 
These integrals can be done exactly, but for later comparison
we will use a saddlepoint method. The saddle points for 
$\alpha_{\w\w'}$ and $\beta_{\w\w'}$ lie respectively at $u= \bar{u}_\pm$
with 
\be
\bar{u}_\pm =  v_0 - {1\over\k} \log(\pm {\w'\over \w})
\ee
The saddlepoint $u = \bar{u}_+$ that contributes to $\alpha_{\w\w'}$ 
corresponds to the physical reflection time,
which is uniquely determined for given in and out-frequency via the Doppler
relation 
\be
\w'=\w \frac{1+{\rm v}}{1-{\rm v}}
\ee
with v the velocity of the mirror given in (\ref{velocity}).
The saddlepoint $u = \bar{u}_-$ that contributes to $\beta_{\w\w'}$, on the
other hand, has an imaginary part equal to $\pi/\k$, and thus corresponds 
to the virtual reflection process that gives rise to the particle creation
phenomenon. These out-going reflection times correspond to 
ingoing times $\bar{v}_\pm$ with
\be
\bar{v}_\pm = v_0 \mp {\w'\over \k \w}
\ee
Hence for the Bogolyubov coefficients we find
$$
\alpha_{\w\w'}\approx 
e^{i\w'\bar{u}_+ - i\w \bar{v}_+}
\approx e^{-\frac{i\w'}{\k}\ln(\frac{\w'}{\w})-iv_0(\w-\w')+\frac{i\w'}{\k}}
$$
\be \beta_{\w\w'}\approx e^{i\w'\bar{u}_- + i\w \bar{v}_-}
\approx e^{-\frac{i\w'}{\k}\ln(-\frac{\w'}{\w})+iv_0(\w+\w')+\frac{i\w'}{\k}}.
\ee 
The out-going radiation spectrum is given by 
$$ 
F(\w')=
\sum_{\w}|\beta_{\w\w'}|^2
$$ 
So, knowing that
$\sum_{\w}(|\alpha_{\w\w'}|^2-|\beta_{\w\w'}|^2)=1$ and that the ratio
$|\alpha_{\w\w'}|^2/|\beta_{\w\w'}|^2$ is independent of $\w$, we get
\be 
F(\w')=\frac{1}{
|\frac{\alpha_{\w\w'}}{\beta_{\w\w'}}|^2-1}=
\frac{1}{e^{2\pi\w'/\k}-1} 
\ee 
which is precisely the Planck spectrum
for thermal (bosonic) radiation with temperature $k_B
T=\frac{\k}{2\pi}$, \cite{dav&ful}.

\newsection{Backreaction on the Mirror}

The typical energy of the out-going particle radiation
is of the order of $\k$ (times $\hbar$). Note however that, given the
outgoing energy $\w'$, the energy $\w$ of the incoming wave will become
exponentially large at late times. This effect 
arises from the fact that the mirror remains accelerating for all
times. However, energy conservation (which should also apply in
the virtual scattering process used in computing the Bogolyubov coefficients) 
tells us that the total energy at early
times can not be more then the total energy of the combined system
at late times. It is
easy to see that this implies that the above computation of the Bogolyubov
coefficients breaks down at relatively early times in case the mirror
has a finite mass.

\begin{figure}[t]
\hspace{1in}
\psfig{figure=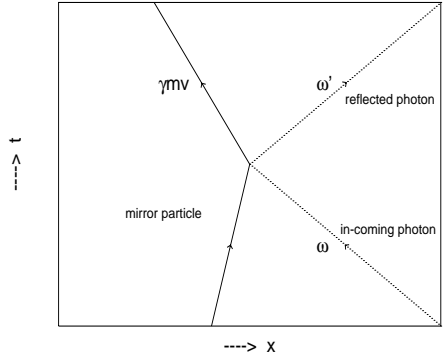,width=2.35in,height=1.85in}
\vspace{1.1in}
\caption{{\footnotesize The reflection of a photon off a mirror particle with
finite mass.}}
\end{figure}

We will therefore now take the mirror mass to be {\em finite} and impose
energy conservation upon reflection.  Consider (in 1+1 dimensions) a
mirror particle with mass $m$ and energy $\gamma$(v) $m$ which collides
elastically with a photon with energy $\w$. The photon reemerges with
a energy $\w'$. We again consider the reflection process 
and we want to express the energy $\w'$ in
terms of $\w$ and the mirror velocity v just {\em after} the collision. 
Classically, one finds: 
\be 
\label{botsing}
\w'=\w\frac{1}{\delta^2-\frac{2\w}{m}\delta} 
\ee 
with $\delta$ the Doppler factor
$$
\delta=\sqrt{\frac{1-{\rm v}}{1+{\rm v}}}.
$$

We now wish to consider the same classical mirror trajectory as in the
previous section. We imagine therefore that there is some external 
force that acts on the mirror particle, which in the absence of other
forces (such as those due to possible (virtual) collisions with the
photons) will keep it in the given trajectory (\ref{trajectory}).
More specifically, in determining the relevant classical WKB-trajectory,
we keep the mirror trajectory 
{\it after} the collision with the photon fixed, so that it always
remains asymptotic to $v=v_0$.\footnote{The reason for this choice is
that in the calculation of the out-going spectrum of radiation, one
needs to investigate the action of an out-going annihilation operator 
on the final state. While this out-going mode is causally connected 
with the early part of the mirror trajectory, it is space-like separated
from the late asymptotic trajectory. In other words, causality tells
us that the mode {\it commutes} with the operators that measure the 
asymptotic mirror trajectory. (Note that this argument does in fact not 
hold in the black hole situation. In that case, the asymptotic time $v_0$ 
is dynamically determined in terms of the infalling matter forming the 
black hole, and therefore causally connected to the out-going radiation.
See \cite{ch&ver,kiem&ver}).} 
Hence the Doppler factor behaves in 
the out-going time $u$ as before
$$
\delta\approx e^{{\k\over 2} (u - v_0)} \approx {1\over \sqrt{\k(v_0-v)}}
$$ 
Thus applying the general formula (\ref{botsing}), we get the following 
relation between the in- and out-going energies in terms of the out-going 
time $u$
\be
\label{omegu}
\w'\approx \frac{\w}{ e^{\k (u - v_0)}-\frac{2\w}{m}e^{{\k\over 2} (u - v_0)}}
\ee
and in terms of the incoming time we have
\be
\label{omegv}
\w\approx \frac{\w'}{{\k (v_0 - v)}+ \frac{2\w'}{m}\sqrt{\k (v_0 - v)}}
\ee
Note that we get the old relation back in the limit $m\to\infty$. For
a finite mirror mass, however, we notice in particular that the incoming 
energy always remains smaller than the kinetic energy of the mirror 
just after the collision. If we keep $\w'$ fixed, then for late times 
$\w$ approaches this maximal value 
\be 
\qquad \qquad 
\w \approx \frac{m}{2} e^{{\k\over 2} (u - v_0)} \approx \gamma m \qquad \qquad
\mbox{for $u \rightarrow \infty$}
\ee 
since $\gamma\equiv \delta/(1-{\rm v}) \approx \delta/2$ for late times.

\newcommand{\x}{{\rm x}}

We would again like to 
use a combination of a WKB approximation for determining the relation
between the in- and out-going wave packets and a saddlepoint approximation 
for the integrals in the expressions of the (corrected) 
Bogolyubov coefficients $\tilde{\alpha}_{\w\w'}$ and $\tilde{\beta}_{\w\w'}$.
Following the same steps as in section 2, we can again express these 
coefficients in terms of the reflection points 
($\tilde{u}_\pm$, $\tilde{v}_\pm$) as
$$
\tilde{\alpha}_{\w\w'}\approx 
e^{i\w'\tilde{u}_+ - i\w \tilde{v}_+}
$$
\be 
\tilde\beta_{\w\w'}\approx e^{i\w'\tilde{u}_- + i\w \tilde{v}_-}.
\ee 
That is, the coefficients are just equal to the ``phase jump'' between the
incoming and outgoing wave at the reflection point.

Using equation  (\ref{omegu}), we find the new saddlepoints 
$\tilde{u}_\pm$ at 
\be 
\tilde{u}_\pm\! = v_0-{1\over \k} \log({\w'\over \w}) +
{2\over\k}\,\log ( \sqrt{\x^2\pm 1} \, \pm \, \x)
\ee
while from (\ref{omegv}) we obtain
\be 
\tilde{v}_\pm\! = v_0 - {\w'\over \k \w}(\pm 1 
+ 2\x^2 - 2\x \sqrt{\x^2\pm 1})
\ee
with
$$
\x^2 = {\w\w'\over m^2}.
$$
As before, the physical reflection point $(\tilde{u}_+,\tilde{v}_+)$ 
that contributes to
the Bogolyubov coefficient $\tilde{\alpha}_{\w\w'}$ 
is always real, while the saddle-point 
$(\bar{u}_-,\bar{v}_-)$ that 
contributes to $\tilde{\beta}_{\w\w'}$ again has an imaginary part.
This imaginary part, however, is no longer constant, but depends on
the variable $\x$. Note further that the virtual reflection point 
$(\bar{u}_-,\bar{v}_-)$ is related to the real one by changing the
sign of the {\it in-going} frequency $\w$. Indeed, the out-going
radiation is produced because negative energy in-coming modes can 
propagate via virtual trajectories into positive energy out-going
modes.

It is easy to see that, if we keep the out-going frequency $\w'$ constant 
(which is still physically reasonable), the incoming frequency $\w$ will
at late times still grow exponentially in the out-going time, but now with 
half the e-folding time. This implies in particular that at late times 
(that is for $\k(\bar{u}_+\!  -v_0) >\!\! >1$), we have the relation
$$
\tilde{u}_+ \approx {2\over \k } \log {\x^2}  + {\rm const.}
$$
between the parameter $\x$ and the reflection time $\tilde{u}_+$.
In the following we will make use of this relation to read off the
time dependence of the radiation spectrum, given its expression in 
terms of the in and out-going frequencies.

A straightforward calculation now gives the following result 
for the new Bogolyubov coefficients:
\be
\tilde\alpha_{\w\w'}\approx \alpha_{\w\w'} \, \exp{2 i\w'\over\k}
\Bigl[-\x (\sqrt{\x^2+1}-\x)   + \log(\sqrt{\x^2+1} +\x) \Bigr]
\ee
and
\be
\qquad
\tilde{\beta}_{\w\w'}\approx \beta_{\w\w'} \, \exp{2 i\w'\over\k}
\Bigl[-{i\pi\over 2}+\x (\sqrt{\x^2-1}-\x)   + \log(\sqrt{\x^2-1}-\x)  \Bigr]
\ee
where $\alpha_{\w\w'}$ and $\beta_{\w\w'}$ denote the uncorrected
coefficients given in section 2. Note that we indeed get the old result 
back in the limit x$\to 0$, which corresponds to the infinite mass 
limit $m \to \infty$.

\begin{figure}[t]
\hspace{1in}
\psfig{figure=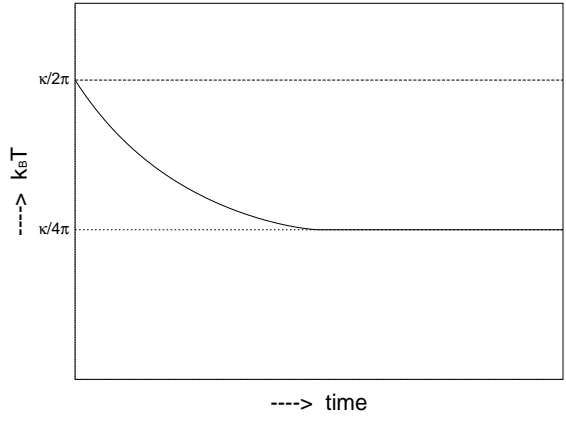,width=3 in,height=2.15in}
\vspace{1.1in}
\caption{{\footnotesize The temperature of the radiation as a function of
the out-going time $u = {2\over \kappa} \log x^2$.}}
\end{figure}

For the ratio of the absolute values of the new coefficients we find
\be
{|\tilde\alpha_{\w\w'}|^2\over |\tilde\beta_{\w\w'}|^2} \approx 
\exp\Bigr[{4\w' \over \k} R(\x) \Bigr]
\ee
with 
\begin{eqnarray}
\qquad \qquad  \qquad \qquad
R (\x) \! \! & = &\! \! \pi -\arccos \x
 + \x \sqrt{1-\x^2} \hspace{10pt} \quad \mbox{ for x} < 1 \nonumber\\[2.5mm]
& = & {\pi}
\quad \qquad \qquad \qquad \qquad \qquad 
 \mbox{for x} > 1.
\end{eqnarray}
We now define the out-going spectrum $F(\w',\x)$ as a function of the
time parameter $\x$ as
\be
F(\w',\x) = {1\over 
{|\alpha_{\w\w'}|^2\over |\beta_{\w\w'}|^2} -1}= \frac{1}{e^{\w'/k_B T(\x)}-1}
\ee
where we identified the time-dependent temperature as 
\be
k_B T(\x) = {\kappa \over 4 R(\x)}.
\ee
The temperature as a function of the outgoing time $\bar{u}$ is plotted in
figure 3.
We read off that after a relatively short amount of time of the
order of 
\be
\bar{u} \approx v_0 + {2\over \k} \log({2m\over \k })
\ee
(recall that the typical frequency of the out-going radiation
is of the order of $\k$) the mirror radiates a constant 
flux of thermal radiation at a 
temperature $k_B T=\frac{\k}{4\pi}$.

In conclusion, we see that introducing a reflecting boundary with 
{\em finite} mass
and imposing energy conservation upon reflection, leads to back
reaction effects with dramatic consequences: the temperature of the
mirror is half of the original result. This result could have been
anticipated from equation (\ref{omegv}). For late
times, i.e. in the limit $v \rightarrow v_0$, the second term in the
denominator determines the corrected Doppler relation between the
in- and out-frequencies. In other words, the magnitude of the ingoing
frequency for given $\w'$ grows linearly with the Doppler factor $\delta$,   
instead of quadratically.

\newsection{Dilaton gravity}

Over the past few years, much effort has gone into the study
of two-dimensional dilaton gravity models of black holes \cite{dilaton}.
We will now make a few comments about
the application of the present WKB method to this 
model. We will not present any calculation, but refer for more
details to \cite{ch&ver,dilaton}.

In the most concrete formulation of dilaton gravity, one imposes reflecting
boundary conditions at some (arbitrarily chosen) critical value of the
dilaton field. If one further chooses the matter fields 
to be massless scalars, one obtains a concrete correspondence with
a dynamical moving mirror model. In this way, the model still shares
many properties with the s-wave reduction
of standard Einstein gravity,  where the boundary corresponds to 
the $r=0$ point. In particular, one finds that all
incoming particle wave of energy below some critical value $\w_{crit}$
will reflect back to infinity, provided it does not disappear behind
the horizon of an already existing black hole. For larger incoming energies
than the critical value $\w_{crit}$, the particle wave will never reflect 
back, but always lead to black hole formation.

\begin{figure}[t]
\psfig{figure=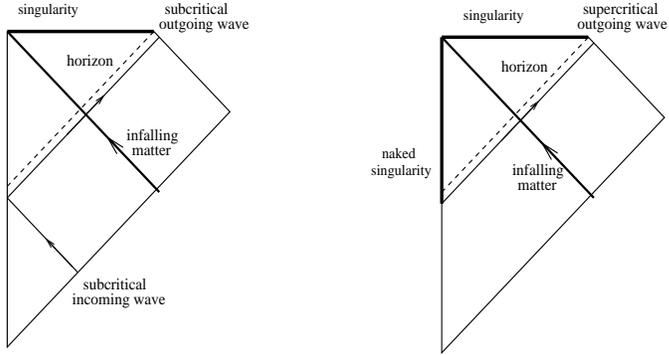,width=3.50 in,height=1.85 in}
\vspace{1.1in}
\caption{{\footnotesize The classical space-times corresponding to 
sub-critical and super-critical out-going waves, resp. The classical
WKB-trajectories that should give the (corrected) Bogolyubov coefficients 
for super-critical out-going waves involve space-times with naked 
singularities, and therefore do not lead to a definite prediction
for the out-going spectrum.}}
\end{figure}

In the sub-critical regime one can calculate the relation 
between the in- and out-going frequencies by solving for the corresponding
classical geometry, see \cite{ch&ver}. 
In the background geometry
of a black hole of mass $M$, we send a signal with frequency $\w'$ 
backwards in time from an outgoing time $\overline{u}$. 
It will bounce off
the boundary, and produce an incoming signal at past infinity.
The relation between the initial
frequency and final frequency can be written for late times ${\overline{u}}$ 
as
\be
\label{relation}
\overline{u} =  v_0 -{1\over \lambda} \log({\w'\over \w})  ,
\ee
with $\lambda$ the dilaton gravity ``cosmological constant'', which
sets the scale of the surface gravity at the black hole horizon.
The time $v_0$ is roughly the black hole formation time.
The reflection off the boundary takes place at an ingoing time $\overline{v}$
given by
\be
\label{time}
\overline{v} = v_0 + {1\over \lambda} 
\log\Bigl({ \w_{crit} \; - \w\over \lambda}\Bigr).
\ee
Somewhat surprisingly, we see in (\ref{relation}) that the backreaction 
of the geometry does not modify the old linearized relation between 
the in- and out-going frequencies. Hence,
if we would use this relation to compute the Bogolyubov coefficients and the 
out-going spectrum, we find no interesting corrections to the out-going
spectrum. We have for $\w < \w_{crit}$
$$
\tilde{\alpha}_{\w\w'}\approx 
e^{i\w'\overline{u}_+ - i\w \overline{v}_+} \approx 
e^{i v_0(\w'-\w) -{i \w' \over \lambda} \log({\w'\over \w})  
- {i\w \over \lambda} 
\log({ (\w_{crit} \; - \w)/ \lambda})}.
$$
\be 
\tilde\beta_{\w\w'}\approx e^{i\w'\tilde{u}_- + i\w \tilde{v}_-}\approx 
e^{-{\pi \w'\over \lambda} + i v_0(\w'+\w) -{i \w' \over \lambda} 
\log({\w'\over \w})  
+ {i\w \over \lambda} 
\log({ (\w_{crit} \; + \w)/ \lambda})}
\ee 
So that
\be
\frac{|\alpha_{\w\w'}|^2}{|\beta_{\w\w'}|^2} = e^{{2\pi \w' \over \lambda}}
\ee
which seems to indicate that the out-going thermal spectrum receives no 
corrections whatsoever.

However, the result (\ref{relation}) for the classical reflection
time is only valid in the sub-critical regime. As soon as 
\be
\label{regime}
\w > \w_{crit}  
\ee
we can no longer use the above formulas, because there no longer exists
a classical trajectory that relates an outgoing wave of this frequency
to a regular incoming wave. Indeed, as seen from (\ref{time}), there
is no longer a physical reflection time. Instead, the only classical 
solutions that contain out-going particles in the supercritical 
regime (\ref{regime}) 
are solutions that contain either white holes or naked singularities. 
In neither case, however, we know of good physical principles that provide 
concrete initial conditions for the out-going state. This super-critical
regime (\ref{regime}) is reached very quickly, since in terms of the out-going
frequency $\w'$ the inequality reads
\be
\w'> \w_{crit} e^{-\lambda(\overline{u}-v_0)}.
\ee
Thus we are forced to conclude that the 
WKB method 
breaks down when applied to two-dimensional dilaton gravity.\\[5mm]

{\noindent \sc Acknowledgements}

\noindent
This research is partly supported by a Pionier Fellowship of NWO, a
Fellowship of the Royal Dutch Academy of Sciences (K.N.A.W.), the
Packard Foundation and the A.P. Sloan Foundation.\\

\end{document}